\documentclass[11pt]{article}
\newcommand{\arrowschem}[2]{\raisebox{-2ex}%
	{$\stackrel{\stackrel{\displaystyle#1}{\longrightarrow}}%
	{\stackrel{\longleftarrow}{#2}}$}}
\newcommand{\arrowchem}[1]{{\stackrel{\displaystyle#1}{\longrightarrow}}}

\usepackage{amssymb}
\begin{document}

\centerline{\bf \Large Uniqueness of steady states for a certain chemical reaction}

\medskip
\centerline{Liming Wang and Eduardo Sontag}
\centerline{Department of Mathematics, Rutgers University}

\bigskip

In \cite{arkin}, Samoilov, Plyasunov, and Arkin provide an example of a
chemical reaction whose full stochastic (Master Equation) model exhibits
bistable behavior, but for which the deterministic (mean field) version has a
unique steady state.

The reaction that they provide consists of an enzymatic futile mechanism
driven by a second reaction which induces ``deterministic noise'' on the
concentration of the forward enzyme (through a somewhat artificial
activation and deactivation of this enzyme).
The model is as follows:
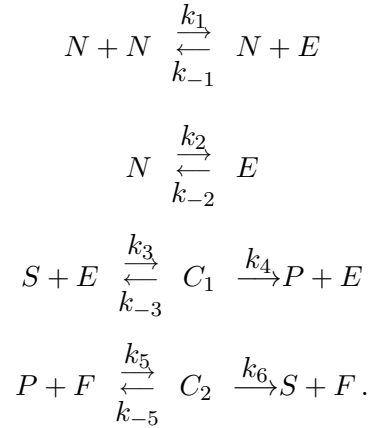
\[
N+N \ \ \arrowschem{k_1}{k_{-1}} \ \ N+E
\]
\[
N\ \ \arrowschem{k_2}{k_{-2}} \ \ E
\]
\[
S+E \ \ \arrowschem{k_3}{k_{-3}} \ \ C_1 \ \ \arrowchem{k_4} P+E
\]
\[
P+F \ \ \arrowschem{k_5}{k_{-5}} \ \ C_2 \ \ \arrowchem{k_6} S+F\,.
\]

Actually, \cite{arkin} does not prove mathematically that this reaction's
deterministic model has a single-steady state property, but shows numerically
that, for a particular value of the kinetic constants $k_i$, a unique steady
state (subject to stoichiometric constraints) exists.  In this short note, we
provide a proof of uniqueness valid for all possible parameter values.

We use lower case letters $n,e,s,c_1,p,c_2,f$ to denote the concentrations of
the corresponding chemicals, as functions of $t$.  The differential equations
are, then, as follows: 
\begin{eqnarray*}
n' &=& -k_1n^2+k_{-1}ne-k_2n+k_{-2}e \\
e' &=& -k_3se+k_{-3}c_1+k_4c_1+k_1n^2-k_{-1}ne+k_2n-k_{-2}e \\
s' &=& -k_3se+k_{-3}c_1+k_6c_2 \\
c_1'&=& k_3se-k_{-3}c_1-k_4c_1 \\
p' &=& k_4c_1-k_5pf+k_{-5}c_2 \\
c_2' &=& k_5pf-k_{-5}c_2-k_6c_2 \\
f'&=& -k_5pf+k_{-5}c_2+k_6c_2\,.
\end{eqnarray*}
Observe that we have the following conservation laws:
\[
e+n+c_1\equiv\alpha\,,\quad
f+c_2\equiv\beta\,,\quad
s+c_1+c_2+p\equiv\gamma\,.
\]

\emph{Lemma 1.}
For each positive $\alpha,\beta,\gamma$, there is a unique (positive) steady
state, subject to the conservation laws.

\emph{Proof.}
Existence follows from the Brower fixed point theorem, since the reduced
system evolves on a compact convex set (intersection of the positive orthant
and the affine subspace given by the stoichiometry class).

We now fix one stoichiometry class and prove uniqueness.
Let $\bar n,\bar e,\bar s,\bar c_1,\bar p,\bar c_2,\bar f$ be any steady state.

From $dn/dt=0$, we obtain that:
\[
\bar e=\frac{k_1 {\bar n}^2+k_2 \bar n}{k_{-1} \bar n+k_{-2}} \,.
\]
From $dc_1/dt=0$, we have:
\[
\bar s=\frac{(k_{-3}+k_{4}) \bar c_1}{k_3 \bar e}\,.
\]
Solving $dc_2/dt=0$ for $p$ and then substituting $f=\beta-c_2$ gives:
\[
\bar p=\frac{(k_{-5}+k_6)\bar c_2}{k_5(\beta-\bar c_2)} \,.
\]
Finally, solving $d(p-f)/dt=0$ with respect to $c_2$ gives:
\[
\bar c_2=\frac{k_4}{k_6}\bar c_1\,.
\]
The derivative of $\bar e$ with respect to $\bar n$ is:
\[
\frac{k_1k_{-1}{\bar n}^2+2k_1k_{-2}\bar n+k_2k_{-2}}{(k_{-2}+k_{-1}\bar n)^2}>0,
\]
and therefore $\bar e$ is strictly increasing on $\bar n$. 

Since $\bar c_1=\alpha - (\bar e+\bar n)$, it follows that $\bar c_1$ is
strictly decreasing on $\bar n$. Therefore $\bar c_2$, $\bar s$, and $\bar p$
are also strictly decreasing on $\bar n$. 

Let $f(\bar n) = \bar s+\bar c_1+ \bar c_2+\bar p$.
Then, $f$ is also decreasing function.

Thus, $\bar n = f^{-1}(\gamma)$ is uniquely defined, and, since 
all coordinates are functions of $\bar n$, it follows that the
steady state is unique, too.

\end{document}